# Negative-contrast neurography: Imaging the extracranial facial nerve and its branches using contrast-enhanced variable flip angle turbo spin echo MRI

**Authors & Affiliations**

Timothy JP Bray[1,2] MBBChir PhD

Emma A Lim[2,4] MBChB BSc

Susan Jawad[2] MBBS BSc

Sumandeep Kaur[2] BSc PgCERT (MMR)

Sofia Otero[2] BM BCh BA

Timothy J Beale[2] MBBS

Mark McGurk[5] MD

Alan Bainbridge[1,3] PhD

[†]Margaret A Hall-Craggs[1,2] MBBS MD

[†]Simon Morley[2] BM BCh MA

[1]Centre for Medical Imaging, University College London, London, United Kingdom

[2]Imaging Department, University College London Hospital, London, United Kingdom

[3]Department of Medical Physics, University College London Hospitals, London, United Kingdom,

[4]Department of Radiology, National Hospital for Neurology and Neurosurgery, Queen Square, London, United Kingdom, [5]Head and Neck Academic Centre, University College London, London, United Kingdom

[†]Joint senior authorship

**Corresponding Author**

Dr Timothy JP Bray MBBChir PhD

Address: Imaging Department, University College Hospital, 235 Euston Road, Bloomsbury, London NW1 2BU, UK

Email: t.bray@ucl.ac.uk

Twitter: @TJP_Bray

Telephone: 0207 679 8156

**Running Title:**

*Negative contrast neurography*

**Abstract**

**Background and Purpose:** Various 'positive-contrast' neurographic methods have been investigated for imaging the extracranial course of the facial nerve. However, nerve visibility can be inconsistent with these sequences and may depend on the composition of the parotid gland, limiting consistent identification. To address this, we describe and evaluate a 'negative-contrast' method for imaging of the extracranial facial nerve using three-dimensional variable flip angle turbo spin echo (VFA-TSE) imaging. We investigate strategies for further optimization, including parotid-specific VFA-TSE optimization and the use of gadolinium-based contrast agent (GBCA).

**Materials and Methods:** 6 healthy volunteers and 10 patients with parotid tumors underwent VFA-TSE and double echo steady state (DESS) imaging of the extracranial facial nerve at 3T. The main trunk, divisions and branches of the extracranial facial nerve were manually segmented by three radiologists, enabling CNR and Hausdorff distance computation and confidence scoring. CNR, Hausdorff distance and confidence scores were compared between sequences and between pre- and post-contrast imaging to evaluate the effect of GBCA.

**Results:** CNR, Hausdorff distances and confidence scores were superior for VFA-TSE compared to DESS imaging. GBCA administration produced a further increase in CNR of nerve against parotid and improved differentiation of nerve from tumor.

**Conclusion:** Imaging of the extracranial facial nerve with VFA-TSE depicts the nerve as a low signal structure ('black nerve') against the high signal parotid parenchyma ('white parotid') and outperforms positive-contrast DESS imaging in terms of CNR, segmentation consistency and confidence. GBCA further increases negative contrast and improves differentiation of nerve from tumor.

**Keywords**

**Introduction**

Injury to the facial nerve is a common and important complication of parotid surgery. Approximately 23% of patients experience postoperative weakness following parotidectomy when intraoperative facial nerve monitoring (FNM) is used; this figure rises to 35% without the use of FNM (1). The incidence of parotid gland tumors is approximately 5 per 100,000 individuals per year (approximately 3,000 per year in the UK and 15,000 per year in the USA) (2), and it can be inferred that thousands of patients might undergo facial nerve injury each year in the Western world. Injury to the facial nerve can also occur during maxillofacial, oral and cosmetic surgery (3). Concerns around facial nerve injury also dictate that deep head and neck lesions in the retromandibular/parotid region are rarely biopsied and that cryoablation of head and neck tumors is rarely used (4,5).

As a result of these issues, efforts have been made to develop dedicated MRI techniques to depict the course of the facial nerve, to help plan surgery and thus to minimize the risk of facial nerve injury. Unfortunately, conventional MRI sequences [such as the widely used two-dimensional turbo spin echo (TSE) sequences] are limited in terms of spatial resolution (both in-plane resolution and slice thickness), meaning that the facial nerve (the main trunk and particularly its branches) are difficult to identify. As a result of these limitations, specialized 'neurographic' imaging methods have been investigated for intraparotid facial nerve imaging, with promising initial results (6–9). The majority of these are 'positive-contrast' methods aiming to extract high signal from the nerves of interest while suppressing signal from all other structures using a combination of relaxation- and diffusion-weighting (6–10). For example, Qin et al. showed that the intraparotid facial nerve could be depicted using double-echo steady state (DESS) sequences and described optimization of acquisition parameters to maximize contrast between the nerve and parotid tissue (6). Fujii et al. subsequently showed this method enabled direct visualization of the intraparotid facial nerve in relation to parotid tumors, with higher localization accuracy than indirect methods for inferring nerve position (7). However, no comparator sequences were included in either study. Similarly, Guenette et al. and Chu et al. showed that the constructive interference in steady state (CISS) and diffusion-weighted reverse fast imaging with steady-state precession (PSIF) sequences respectively could display the main trunk and divisions of the nerve, but neither study compared against other sequences (8,9). However, a drawback of positive contrast sequences is that the achievable contrast depends on the composition of the background tissue (particularly water content); nerve visibility is therefore variable and – in our experience - often limited in clinical practice. Positive-contrast methods also suppress signal from structures other than the nerve, meaning that anatomical detail needed for surgical planning is lost. As a result of these factors, there is an unmet need for improved neurographic methods which consistently and reliably show the extracranial facial nerve to guide surgical planning.

Here, we propose an alternative approach to imaging of the extracranial nerve based on variable flip angle turbo spin echo (VFA-TSE) imaging. The VFA-TSE sequence is a modified turbo spin echo sequence which uses variable flip angles through the echo train in order to produce an almost constant echo amplitude through the train, thus increasing the number of usable echoes and sharpening the point spread function (11,12). As with standard TSE, k-space lines are acquired at a range of different echo times, and the acquisition is said to have an effective echo time ($TE_{eff}$), which corresponds to the echo time at which the central lines of k-space is acquired.

The flip angle scheme is prescribed based on the $T_1$ and $T_2$ values of the tissue of interest (13); however, the prescribed signal evolution depends relatively weakly on $T_1$ and $T_2$ and is similar for many tissues. For neurographic imaging, VFA-TSE imaging offers several advantages over existing techniques, including (a) the potential for high-resolution 3D imaging with minimal blurring (clearly an advantage when imaging very small structures) and (b) spin echo-based contrast (which avoids the problem of $T_2^*$ shortening due to inhomogeneities created by the mixture of water and fat in the parotid). Although VFA-TSE does not provide nerve-specific positive contrast, we reasoned that - with appropriate optimization and weighting - the nerve might visualized as a low signal structure ('black nerve') against the high signal parotid ('white parotid'), which typically contains a substantial proportion of fat. We also reasoned that this negative contrast might be improved further by the use of parotid-specific optimization of the VFA-TSE sequence, and through the use of gadolinium-based contrast (GBCA), since this might produce enhancement of small vessels and the parotid parenchyma and therefore increase the difference in signal intensity between parotid parenchyma and nerve. Specifically, we hypothesized that $T_1$-weighted high-resolution TSE sequences would offer superior contrast-to-noise ratio (CNR), reader confidence and segmentation consistency compared to DESS imaging and $T_2$-weighted TSE sequences. We also hypothesized that the use of GBCA and parotid-specific optimization of the VFA-TSE sequence would improve CNR.

**Methods**

*Study Overview*

This prospective study was approved by the Queen Square Research Ethics Committee, London (reference 15/LO/1475) and participants gave written informed consent. First, six healthy volunteers aged between 24y8m and 33y5m with no history of parotid tumor and no contraindication to MRI scanning were prospectively recruited and scanned. Each volunteer was imaged with three methods (see details below), without GBCA, for the purposes of method comparison and to determine the best-performing sequence for further optimization. A further volunteer aged 27y9m was scanned to evaluate the effect of parotid-specific relaxation times for optimizing of the VFA-TSE sequence. Following optimisation, ten consecutive patients aged between 38y8m and 53y9m with parotid tumors were scanned with a dedicated facial nerve neurography protocol consisting of pre- and post-contrast VFA-TSE imaging in order to determine the visibility of the nerve in the presence of tumor and to evaluate the effect of GBCA (on both the side with tumor and the side without). Patients were included in the study if they (i) had a known or suspected benign or malignant parotid tumor based on clinical examination and/or previous imaging and (ii) were being considered for resection of the parotid tumor. Patients were excluded if they had a contraindication to MRI scanning or to the administration of GBCA.

*Sequences*

All subjects were imaged on a Siemens Vida 3T scanner with a standard 20-channel head coil (without surface coils) (Siemens, Erlangen, Germany). Each volunteer was imaged with three methods (see Table 1), consisting of two implementations of VFA-TSE, optimized for $T_1$-weighting and $T_2$-weighting respectively, and a double echo steady state (DESS) sequence. The VFA-TSE acquisitions were investigational versions of the Siemens SPACE sequence, with prescribed flip angle schemes based on the relaxation parameters shown in Table 1 (13). The flip angle scheme for the $T_1$-weighted VFA-TSE acquisition was prescribed based on $T_1$=940ms and $T_2$=100ms (13); these are standard values used for brain imaging. To assess the benefit of parotid-specific optimization of the VFA-TSE sequence, a volunteer was scanned with two flip angle schemes – first with $T_1$=940ms and then with $T_1$=500ms to more closely match the value expected in parotid tissue. To assess the benefit of GBCA, all patients with parotid tumors underwent Method 1 (T1w VFA-TSE) acquisitions before and after hand injection of 10ml of gadoteridol 279.3mg/ml. The DESS sequence used spectrally selective excitation of water; other parameters were chosen to approximately match the implementation of Fuji et al. (7).

[Insert Table 1]

*Image Analysis*

For the volunteer study, images from the three sequences were analysed separately and in random order by three radiologists, comprising two head and neck radiology consultants (SM and SJ, with 20 and 6 years of experience in head and neck imaging respectively) and a neuroradiology fellow (EL, with two years of neuroradiology experience) using 3D Slicer (Version 4.10) (14,15). For all images, the facial nerve and its branches, main parotid duct and parotid gland were viewed in three planes and manually segmented. Each structure was assigned a separate numerical segmentation label, such that a segmentation image consisting of 0 entries for no segmentation and the number of the label for the segmented areas was generated. For the facial nerve, labelled segmentations were generated for (1) the main trunk of the extracranial component of the facial nerve (i.e. the segment extending from the stylomastoid foramen to the pes anserinus), (2) the upper division of the nerve and its continuations (the temporal and zygomatic branches) (3) the lower division of the nerve and its continuations (the marginal mandibular and cervical branches) and (4) the buccal branch (see Figure 1). The branches of the facial nerve were only segmented within the confines of the parotid gland; the extraparotid components of these nerves were not evaluated. Further segmentations were generated for the main parotid duct and large branches and for the retromandibular vein, excluding parts of these structures lying outside the boundaries of the parotid gland. A separate segmentation of the parotid was generated for each subject by an additional reader (TJPB, a radiology registrar with one year of head and neck experience and six years of MRI research experience) in order to provide a single, fixed segmentation of the parotid for CNR calculation (since variations in parotid segmentation by the three primary readers might otherwise influence the CNR). This reader also evaluated all images for the presence of motion artifact, and any images with significant motion artifact were excluded from subsequent analysis.

Segmentation masks for each reader were exported into a dedicated in-house analysis module written in MATLAB (TJPB) enabling calculation and tabulation of CNR and Hausdorff distances for each nerve segment (designated by segmentation label), sequence and reader.

[Insert Figure 1]

For each of the nerve segments (i.e. main trunk, upper division and branches, lower division and branches and buccal branch), CNR was calculated for the nerve relative to the background tissue using

$$CNR = \frac{S_{nerve} - S_{background}}{\sqrt{(SD_{nerve}^2 + SD_{background}^2)/2}}$$

where $S_{nerve}$ and $S_{background}$ are the signal intensities for the nerve and background tissue respectively and $SD_{nerve}$ and $SD_{background}$ are the standard deviation values for the nerve and background respectively. In healthy volunteers and on the normal side in patients, the background tissue was the parotid gland; in patients on the tumor side, the background tissue was the tumor. Note that CNR values are negative where the nerve is darker than the background tissue but positive where the nerve is brighter than the background. Note that raw CNR values (including both positive and negative values) rather than absolute values were used for the statistical analysis because the magnitude operation can potentially introduce a bias for methods where the contrast between nerve and parenchyma can be either positive or negative (depending on the composition of the parotid gland); for example, a method with a mean CNR of zero but a high variance between subjects would have a substantial positive CNR if the magnitude was taken.

Euclidean Hausdorff distances between observers 'segmentations were calculated for each nerve segment using a fast method described by Lindblad (16). In the context of segmentation masks, the Hausdorff distance is the maximum distance between any pixel in the first mask and the nearest pixel in the second mask and provides a measure of the similarity or 'matching 'of those masks. All images were also rated in terms of the observer's confidence in the visibility of the each nerve segment, on a three-point scale where 0 is 'not visible', 1 is 'partially visible 'and 2 is 'completely visible'.

*Statistical Analysis*

CNR and Hausdorff distances were compared between sequences and between pre- and post-contrast images using multilevel mixed-effects linear regression analysis including data from all nerve segments (main trunk, upper division and lower division). Specifically, 'sequence 'was used as a categorical predictor variable, CNR or Hausdorff distance were used as outcome variables and patient identification number, nerve segment and observer were used as grouping variables. Confidence scores were first averaged over readers; differences between sequences were evaluated using the Friedman test with Dunn's test for multiple comparisons and the effect of contrast was evaluated using the Wilcoxon matched-pairs signed rank test. Mixed effects analyses were performed using Stata/IC 14.1 for Mac (StataCorp, Lakeway Dr, College Station, Tx); the Friedman and Wilcoxon tests were performed using Prism 7.0e for Mac (GraphPad, Northside Dr, San Diego, Ca).

**Results**

Examples of images from both healthy subjects and patients are shown in Figure 1. Contrast-to-noise ratio measurements for the various imaging sequences in both healthy volunteers and patients are shown in Figure 2. Hausdorff distances are shown in Figure 3, and confidence scores are shown in Figure 4. Figure 5 shows the effect of sequence on CNR and confidence scores separately for each nerve segment. Examples of images acquired with Method 1 before and after GBCA are shown in Figure 6.

[Insert Figures 2-6]

***Sequence Comparison in Volunteers***

Estimated CNR (95% CI) was -1.39 (-1.62 to -1.16) for Method 1, -1.19 (-1.44 to -0.95) for Method 2 and +0.214 (-0.82 to +0.51) for Method 3]; CNR differed significantly for both Method 1 and Method 2 compared to Method 3 (P=0.000 and 0.000 respectively) (Figure 2a).

Estimated Hausdorff distances were 13.6 (4.24 to 22.9) for Method 1, 23.9 (14.1 to 33.8) for Method 2 and 12.8 (0.75 to 24.8) for Method 3; there was no significant different between Methods 1 and 3 (P=0.884) or Methods 2 and 3 (P=0.055) (Figure 3a).

Median (IQR) confidence scores were 1.33 (0.67 to 2) for Method 1, 1.67 (1 to 2) for Method 2 and 1.33 (0.67 to 2) for Method 3; confidence did not differ significantly for either Methods 1 and 3 (P=0.24) or methods 2 and 3 (P=0.09) (Figure 4a).

*Optimization of VFA-TSE Parameters*

For the optimization experiment, there was no change in confidence or CNR depending on the $T_1$ value used for sequence optimization, i.e. parotid-specific optimization produced no benefit compared to standard optimization.

*Nerve Visibility in Patients and Effect of GBCA*

Figure 6 shows examples of pre- and post-GBCA images acquired with Method 1 in patients (both on the side without tumor and on the tumor side). Both pre- and post-contrast images were adequate in 7/10 subjects; three subjects were excluded due to the presence of motion artifact on one or both sets of images.

On the normal side (without tumor), estimated CNR was significantly greater for the post contrast images [-1.91 (-2.13 to -1.69)] than for the pre-contrast images [-1.61 (-1.83 to -1.39)] (P=0.002) (Figure 2b). Hausdorff distances were significantly lower – indicating greater segmentation similarity between observers - for the post-contrast images [24.8 (5.63 to 44.1)] than for the pre-contrast images [20.8 (1.56 to 40.0)] (P=0.001) (Figure 3b). Median (IQR) confidence scores were 1.67 (1.67 to 2) without contrast and 2 (1.67 to 2) with GBCA; confidence was significantly higher on the post-contrast images (P=0.0166) (Figure 4b).

On the tumor side, estimated CNR was significantly greater for the post contrast images [-0.96 (-1.43 to -0.049)] than for the pre-contrast images [+0.17 (-0.30 to +0.63)] (P=0.000) (Figure 2c). Hausdorff distances did not significantly differ for the post-contrast images [14.9 (8.73 to 21.0)] compared to the pre-contrast images [17.8 (11.2 to 24.4)] (P=0.41) (Figure 3c). Median (IQR) confidence scores were 1.67 (0.67 to 2) without contrast and 1.5 (1.33 to 2) with GBCA (P=0.063) (Figure 4c).

**Discussion**

Various methods have been proposed for neurographic imaging of the extracranial nerve, the majority of which rely on positive contrast and show the nerve as a bright structure relative to the low signal parotid gland and the vessels and ducts it contains. Here, we showed that a 'negative contrast' approach to imaging the extracranial facial nerve using VFA-TSE imaging significantly outperformed neurographic imaging using the DESS sequence, a widely-used positive contrast technique. The DESS sequence showed wide variability in the contrast between nerve and parotid, with positive and negative CNR values observed in different subjects, which may reflect variability in the composition of the parotid parenchyma. In contrast, VFA-TSE imaging consistently depicts the nerve as a low-signal structure ('black nerve') against the high-signal parotid parenchyma ('white parotid'), and we showed that the negative contrast between the nerve and parotid might be further enhanced using GBCA. This approach may help to plan surgery in patients with parotid tumors.

Of the methods investigated, we found that Method 1 (i.e. VFA-TSE optimized to provide $T_1$ weighting) offered the best performance across the three metrics (CNR, Hausdorff distance and reader confidence). The negative contrast provided by this method is likely to be underpinned by high fat content in the parotid gland, contributing to shortening of the effective $T_1$ for the tissue and increasing signal on scans optimized for $T_1$ weighting. There may also be a $T_1$-shortening effect of protein in saliva lying within the salivary acini and ducts. Previous studies have also investigated sequences relying on $T_1$ contrast to image the nerve; however, most of these have used gradient-echo based sequences. For example, Dailana et al. used a 3D gradient echo sequence with a short echo time of 4.2ms and could identify the most proximal part of the main trunk of the facial nerve, but the intraparotid main trunk, divisions and branches were not clearly seen (17). From the images in their paper, it is striking that the parotid gland appears much lower signal than with the VFA-TSE method described here; this may be due to the fact that the presence of both fat and water in the parotid parenchyma creates local inhomogeneities, causing shortening of $T_2^*$, whereas $T_2$ is relatively unaffected. Similarly, Tsang et al. compared two gradient-echo based sequences with $T_1$-weighting and were unable to see the divisions or branches of the extracranial facial nerve with either sequence; again the images in the paper show the parotid as only marginally higher signal than the nerve (in comparison to the VFA-TSE technique) (18). A similar phenomenon is seen in the images in the work by Takahashi et al. (19), whilst Kraff et al. also failed to consistently identify the extracranial facial nerve at 7T using various gradient echo based methods (10). We therefore suggest that the use of a spin-echo based technique may be an important component of the negative contrast observed here.

Amongst the spin echo sequences, VFA-TSE is a logical choice due to its ability to deliver high resolution with minimal blurring.

Importantly, we found that imaging after GBCA administration further increased the negative contrast between nerve and parotid and also increased reader confidence. In addition to causing enhancement of the parotid parenchyma itself, the presence of GBCA can produce enhancement of small vessels and reduce their conspicuity, effectively make the parotid gland appear more homogenous. This might reduce the 'distracting 'effect of these vessels (which might be confused for nerve branches). However, it should be noted that – in large vessels – the flow suppression effects of the VFA-TSE sequence might outweigh the effect of GBCA, and thus these structures can still show low signal even on the post-contrast images. In general, there are two potentially opposing effects which might influence vessel enhancement with VFA-TSE sequences: (1) the concentration (and relaxivity) of GBCA in the tissue/vessel and (2) the speed and direction of flow. Whereas increases in GBCA concentration normally produces an increase in signal intensity, VFA-TSE sequences have intrinsic flow suppression properties owing to the large gradient moment arising from the length of the echo train. In our images, it is apparent that large vessels usually show increased signal in the periphery of the vessel lumen on the post-contrast images relative to the pre-contrast images. This likely arises because the velocity of flow is lowest close to the vessel wall, meaning that the $T_1$-shortening effect of gadolinium contributes to increased signal intensity, whereas the signal in the centre of large vessels - where velocity is higher - is suppressed, and therefore any effect of the gadolinium is negated. It seems logical, therefore, that some enhancement can be expected in small vessels in the parotid, although the flow suppression effect will predominate in larger vessels. However, the presence of peripheral enhancement in large vessels might still allow the observer to differentiate these from nerve and thus avoid incorrect labelling of these structures as the facial nerve or its branches.

Our results suggest that the use of GBCA offers a further benefit in patients with tumors, as the tumors tend to be hypointense on the unenhanced VFA-TSE images (and are therefore difficult to differentiate from the nerve) but enhance avidly with contrast, meaning that differentiating between nerve and tumor is much easier on the post-contrast images. By comparison, both tumor and nerve are typically hyperintense on DESS images, making differentiation of these structures difficult [7]. The ability to differentiate nerve from tumor may be of particular clinical value since the operative risk is likely to be highest when these structures lie in very close proximity. We suggest that the acquisition of both pre- and post-contrast images may be optimal, as the pre-contrast images allow better delineation of tumor boundaries whereas the post-contrast images offer more detail on the course of the nerve. Additional acquisitions for relaxometry may also help to distinguish these tissues.

In clinical practice, detailed imaging of the facial nerve and its branches might help to inform the operative approach in patients with parotid tumors and therefore minimize the risk of facial nerve injury. The traditional approach to parotid surgery is to first find the main trunk of the facial nerve before dissecting downwards and excising the tumor; however, if the divisions and/or branches of the nerve can be seen in detail, the approach might be better tailored to individual patients. For example, in patients with tumors that lie completely remote from the nerve, the tumor may be directly excised by extracapsular dissection (where the tumor is excised without formal identification of the nerve), whereas patients with tumors that lie deep to the nerve may need more involved surgery with identification of the nerve and branches [20]. In either approach, prior knowledge of the specific course of the nerve and any divisions/branches that are particularly intimately related to the tumor seems likely to reduce the chance of injury. It may also allow identification of patients in whom injury to the nerve is more likely (due to the proximity of the tumor to the nerve or the specific anatomy) and therefore enable patients to be better informed of the level of risk when consenting to the procedure.

Optimization of methods for visualization and display of the nerve and tumor (suitable for both radiologists and surgeons) may be an interesting avenue for future research. For example, rotating maximum intensity projections (or segmented datasets) may allow clearer depiction of the relationship between the nerve and tumor than traditional MRI reconstructions. The use of augmented reality (AR) – whereby the MRI-derived image could be superimposed on the user's view of the patient - might also offer a means to improve visualization and has the potential benefit that the structures of interest can be viewed in the context of the rest of the patient's anatomy, potentially increasing confidence and surgical precision (21).

***Limitations***

A limitation of this study is that patients with parotid tumors did not undergo DESS acquisitions, meaning that that the comparison between Methods 1-3 was limited to volunteers. This was a pragmatic decision based on scan time and the results of the volunteer experiment – we think it is unlikely that the DESS sequence would offer superior performance in patients with tumors given that it performed poorly in volunteers. A further limitation is that patients typically moved very slightly between acquisitions, meaning that patients were not perfectly aligned. We addressed this issue by generating separate segmentation masks for each acquisition (rather than propagating between acquisitions), meaning that the nerve was accurately segmented (where visible) for each sequence. An alternative approach would have been to register the images between sequences; however, this is not a trivial task as the deformation in the region of the parotid is typically nonrigid. The methodology employed here was therefore the most practical means to mitigate this problem. Similarly, motion artifact within each of the scans makes the nerve more difficult to identify and may have contributed to reduced CNR for some acquisitions. Strategies for minimizing motion (e.g. head stabilisers) or algorithms for motion correction might help to further improve the quality of VFA-TSE imaging in the parotid, and these areas may be useful avenues for further research.

**Conclusion**

Imaging of the extracranial facial nerve with VFA-TSE depicts the nerve as a low signal structure ('black nerve') against the high signal parotid parenchyma ('white parotid') and outperforms the widely used positive-contrast DESS sequence. The use of GBCA further increases CNR and may be of particular value in patients with tumors, as the non-enhancing nerve can be differentiated from enhancing tumor.

**Acknowledgements**

TJPB is supported by an NIHR Clinical Lectureship (CL-2019-18-001). MHC is supported by the National Institute for Health Research (NIHR) Biomedical Research Centre (BRC). This work was undertaken at UCLH/UCL, which receives funding from the UK Department of Health's the NIHR BRC funding scheme. The views expressed in this publication are those of the authors and not necessarily those of the UK Department of Health.

**Tables**

Table 1 – MRI Acquisition Parameters

|  | Method 1 (T$_1$-weighted VFA-TSE) | Method 2 (T$_2$-weighted VFA-TSE) | Method 3 (DESS) |
|---|---|---|---|
| Sequence | 3D variable flip angle turbo spin echo | | Double echo steady state |
| Coil | 20 channel head coil | | |
| Orientation | Sagittal | | |
| Matrix size | 320×320×208 | 288×280×96 | 256x256x40 |
| Resolution | 0.76×0.76x0.80 mm$^3$ | 0.76×0.76x0.80 mm$^3$ | 0.78x0.78x0.81 mm$^3$ |
| TE$_{effective}$ | 25 ms | 147 ms | 4.2 ms |
| Flip angle | Variable: optimized for T$_1$ = 940ms T$_2$ =100ms | 120° | 30° |
| TR | 700 ms | 1200 ms | 11 ms |
| Fat suppression | None | | Water excitation |
| Echo train length | 27 | 70 | - |
| Bandwidth | 625 Hz/Px | 289 Hz/Px | 425 Hz/Px |
| k-space coverage | Full | | |

**Figure Legends**

**Figure 1 - Example images.** Methods (a) 1 (T1w VFA-TSE), (b) 2 (T2w VFA-TSE) and (c) 3 (DESS), and illustrative segmentation labels for (d) individual nerve segments, (e) parotid gland and (f) tumor are shown in the sagittal plane. In (d), the main trunk is segmented in red, the upper division in green and the lower division in blue (numerically these were assigned labels 1, 2 and 3 respectively).

**Figure 2 – Contrast-to-noise ratio.** Raw CNR measurements (including positive and negative values) are shown in (a)-(c); absolute CNR measurements are shown in (d)-(f). Measurements for all nerve segments and all three readers are shown. CNR is compared between sequences (a,d), between pre- and post-contrast images for normal parotid (b,e) and between pre- and post-contrast images in parotid glands with tumor (c,f). Note that CNR scores are negative for Methods 1 and 2 due to the lower signal of the nerve against the parotid. For Method 3, there is substantial variability in CNR, including both positive and negative values, which may be due to variability in the composition of the parotid parenchyma. Note that the raw CNR measurements were used for the statistical analysis as the magnitude operation can introduce a positive bias when the mean CNR is close to 0, as in (a,d). Methods 1-3 correspond to T1w VFA-TSE, T2w VFA-TSE and DESS sequences respectively.

**Figure 3 – Hausdorff distances.** Measurements for all nerve segments and all three readers are shown. Hausdorff distances are compared between sequences (a), between pre- and post-contrast images for normal parotid (b) and between pre- and post-contrast images in parotid glands with tumor (c). Smaller Hausdorff distances

indicate better agreement between readers' segmentations. Methods 1-3 correspond to T1w VFA-TSE, T2w VFA-TSE and DESS sequences respectively.

**Figure 4 – Reader confidence.** Measurements for all nerve segments and all three readers are shown. Confidence scores are compared between sequences (a), between pre- and post-contrast images for normal parotid (b) and between pre- and post-contrast images in parotid glands with tumor (c). Methods 1-3 correspond to T1w VFA-TSE, T2w VFA-TSE and DESS sequences respectively.

**Figure 5 – CNR and confidence by nerve segment.** CNR values by method are shown for the main trunk, upper division and lower division of the facial nerve (a-c). Confidence scores are shown in (d-f). Methods 1-3 correspond to T1w VFA-TSE, T2w VFA-TSE and DESS sequences respectively.

**Figure 6 – Method 1 (T1w VFA-TSE) images before and after GBCA.** The contrast between the parotid gland and nerve slightly increases on the post-GBCA images (a,b). In addition, some small vessels (closed arrowhead) become less conspicuous on the post-GBCA images, reducing the potential for misidentification. In patients with tumor, the tumor (*) and nerve (open arrowhead) have similar signal intensity on the images without GBCA (c); the tumor significantly increases in intensity on the post-contrast images while the nerve remains low signal (d), meaning that CNR increases. This may be of particular importance where the nerve lies in close proximity to the tumor.

**Figures**

**Figure 1**

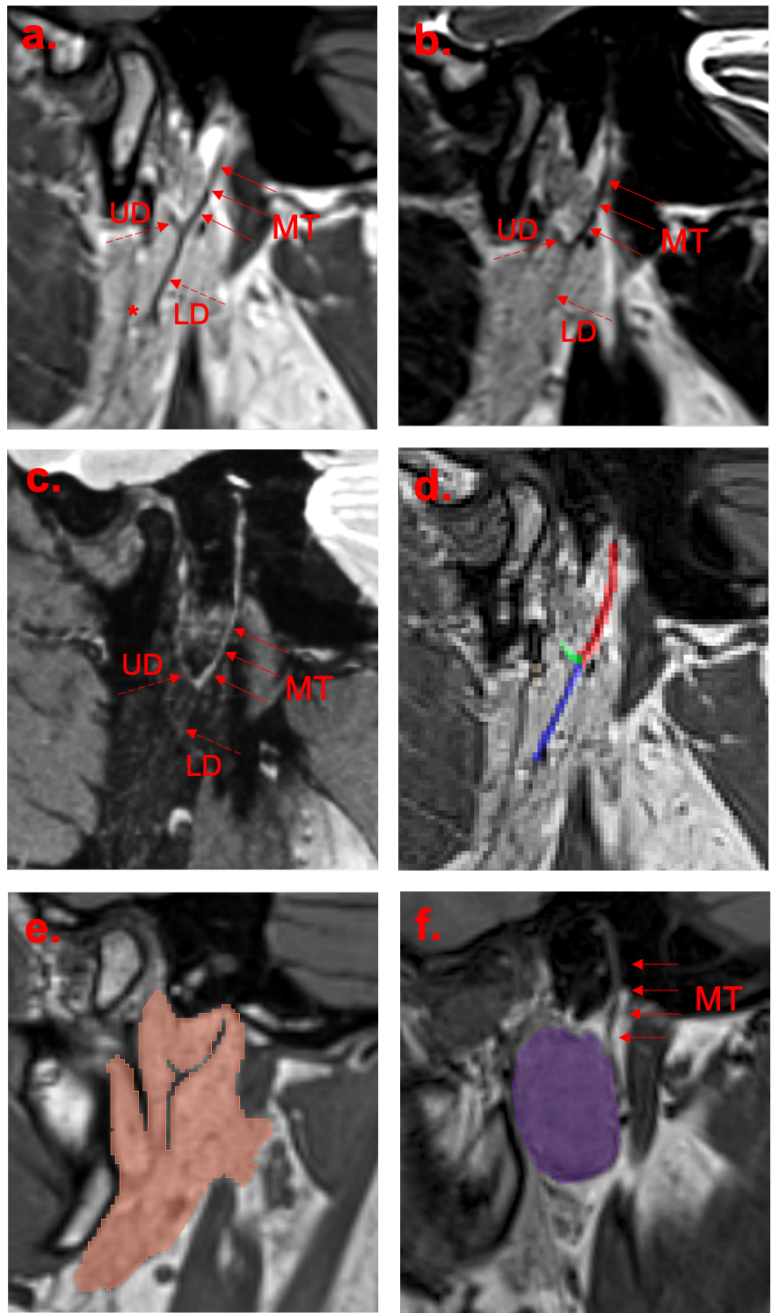

Figure 1 - Example images. Methods (a) 1 (T1w VFA-TSE), (b) 2 (T2w VFA-TSE) and (c) 3 (DESS), and illustrative segmentation labels for (d) individual nerve segments, (e) parotid gland and (f) tumor are shown in the sagittal plane. In (d), the main trunk is segmented in red, the upper division in green and the lower division in blue (numerically these were assigned labels 1, 2 and 3 respectively).

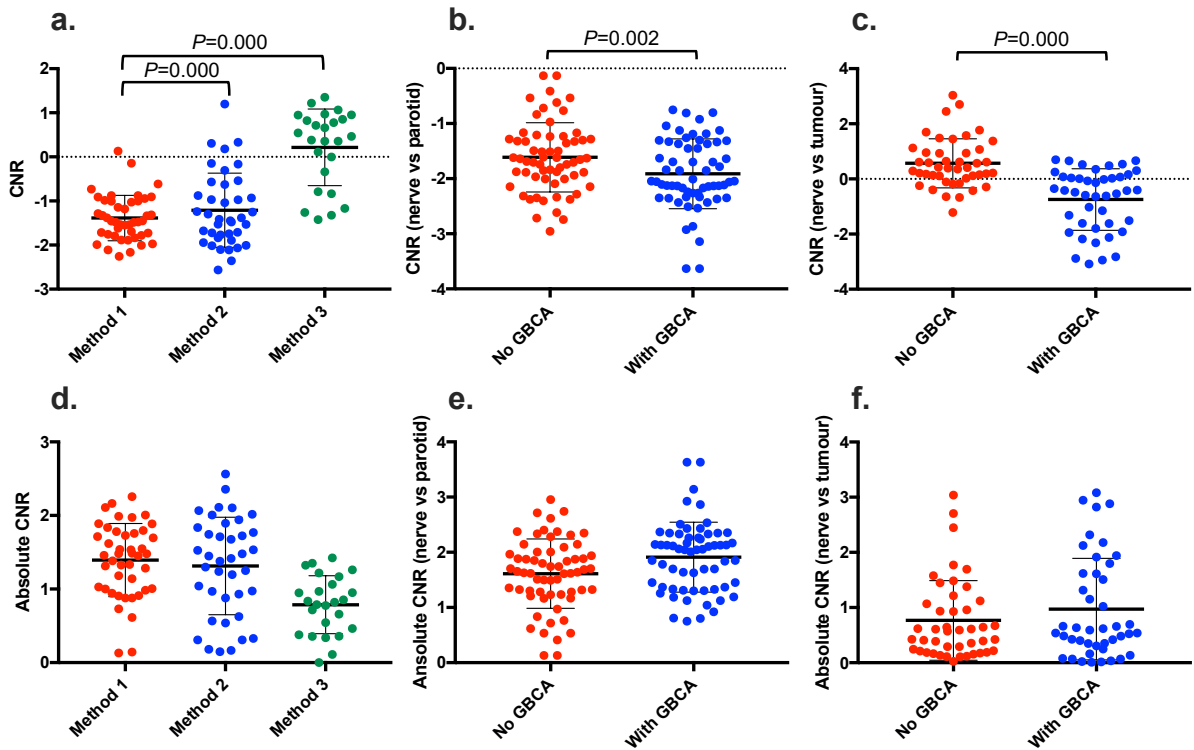

Figure 2 – Contrast-to-noise ratio. Raw CNR measurements (including positive and negative values) are shown in (a)-(c); absolute CNR measurements are shown in (d)-(f). Measurements for all nerve segments and all three readers are shown. CNR is compared between sequences (a,d), between pre- and post-contrast images for normal parotid (b,e) and between pre- and post-contrast images in parotid glands with tumor (c,f). Note that CNR scores are negative for Methods 1 and 2 due to the lower signal of the nerve against the parotid. For Method 3, there is substantial variability in CNR, including both positive and negative values, which may be due to variability in the composition of the parotid parenchyma. Note that the raw CNR measurements were used for the statistical analysis as the magnitude operation can introduce a positive bias when the mean CNR is close to 0, as for Method 3 in (a,d).

Methods 1-3 correspond to T1w VFA-TSE, T2w VFA-TSE and DESS sequences respectively.

**Figure 3**

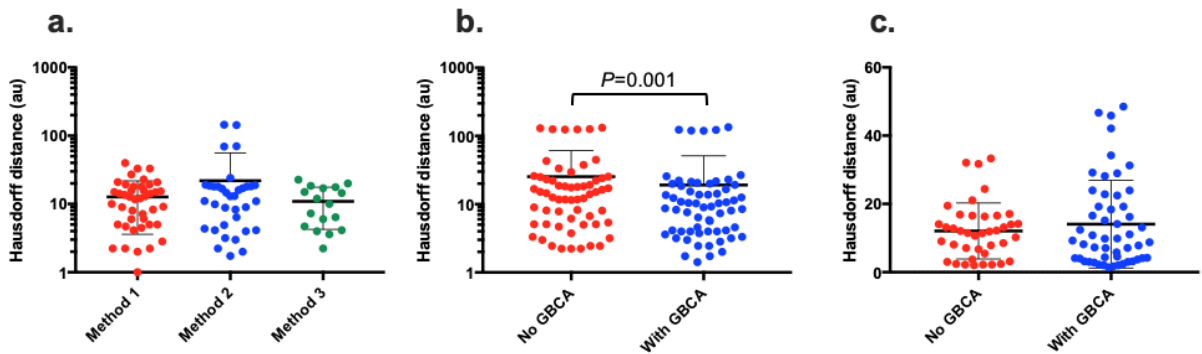

Figure 3 – Hausdorff distances. Measurements for all nerve segments and all three readers are shown. Hausdorff distances are compared between sequences (a), between pre- and post-contrast images for normal parotid (b) and between pre- and post-contrast images in parotid glands with tumor (c). Smaller Hausdorff distances indicate better agreement between readers' segmentations. Methods 1-3 correspond to T1w VFA-TSE, T2w VFA-TSE and DESS sequences respectively.

**Figure 4**

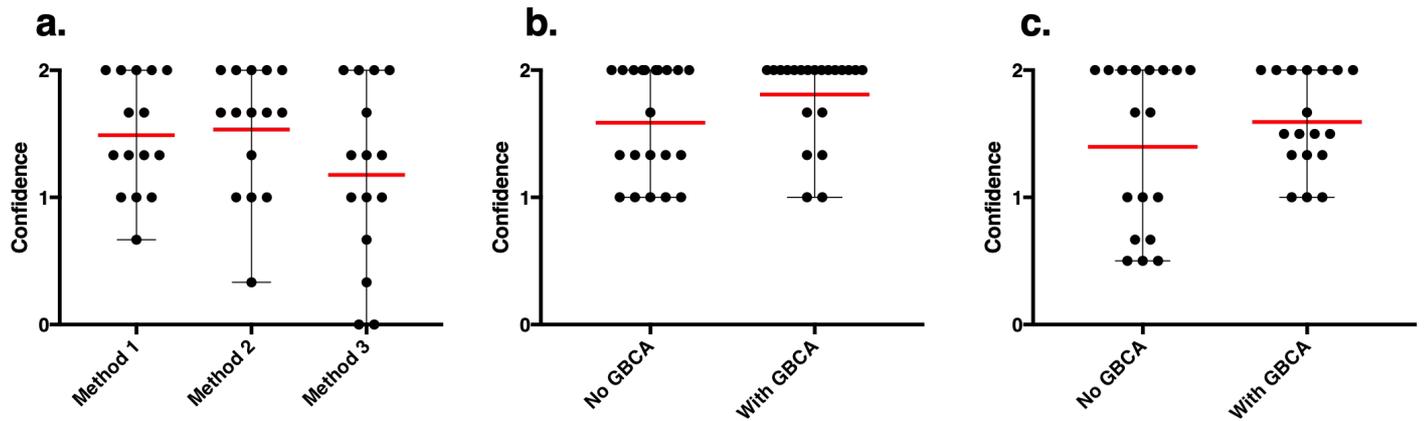

Figure 4 – Reader confidence. Measurements for all nerve segments and all three readers are shown. Confidence scores are compared between sequences (a), between pre- and post-contrast images for normal parotid (b) and between pre- and post-contrast images in parotid glands with tumor (c). Methods 1-3 correspond to T1w VFA-TSE, T2w VFA-TSE and DESS sequences respectively.

**Figure 5**

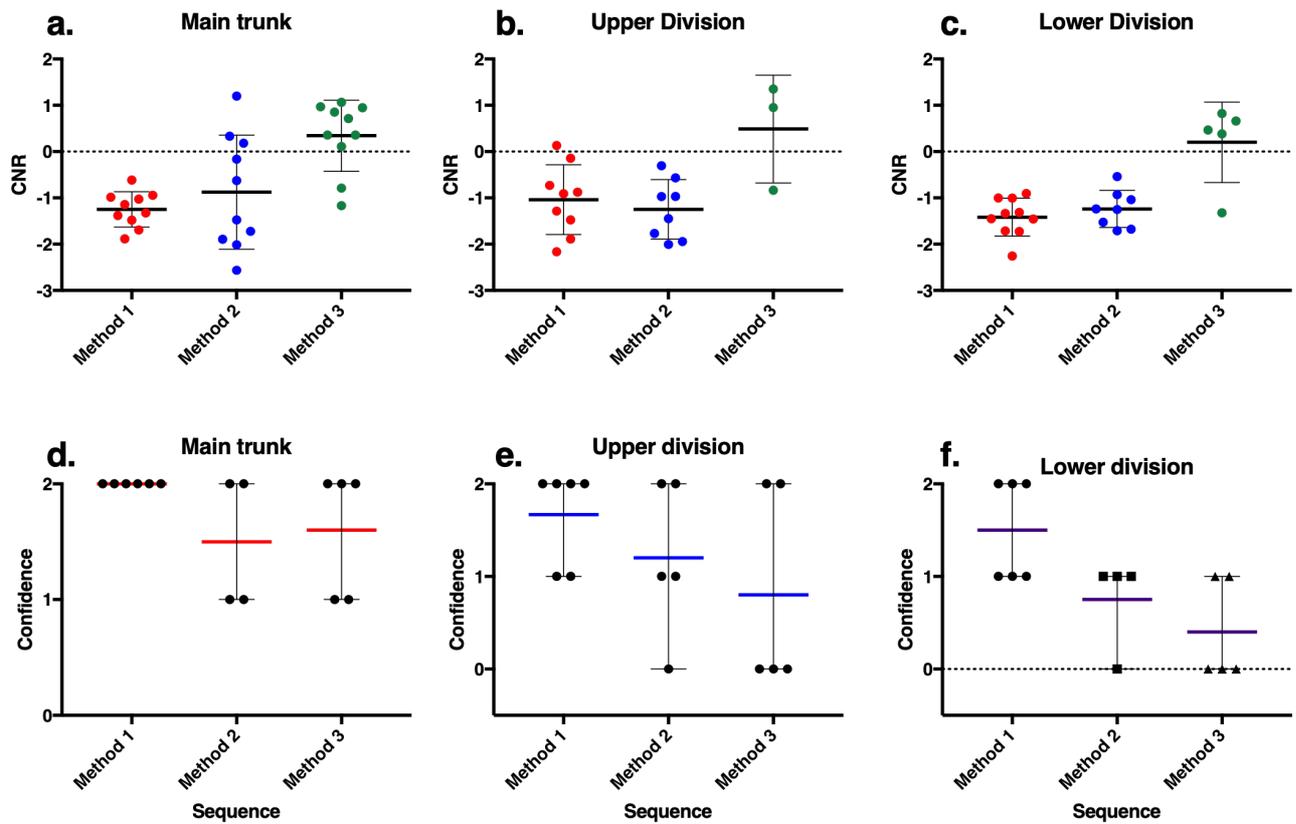

Figure 5 – CNR and confidence by nerve segment. CNR values by method are shown for the main trunk, upper division and lower division of the facial nerve (a-c). Confidence scores are shown in (d-f). Methods 1-3 correspond to T1w VFA-TSE, T2w VFA-TSE and DESS sequences respectively.

**Figure 6**

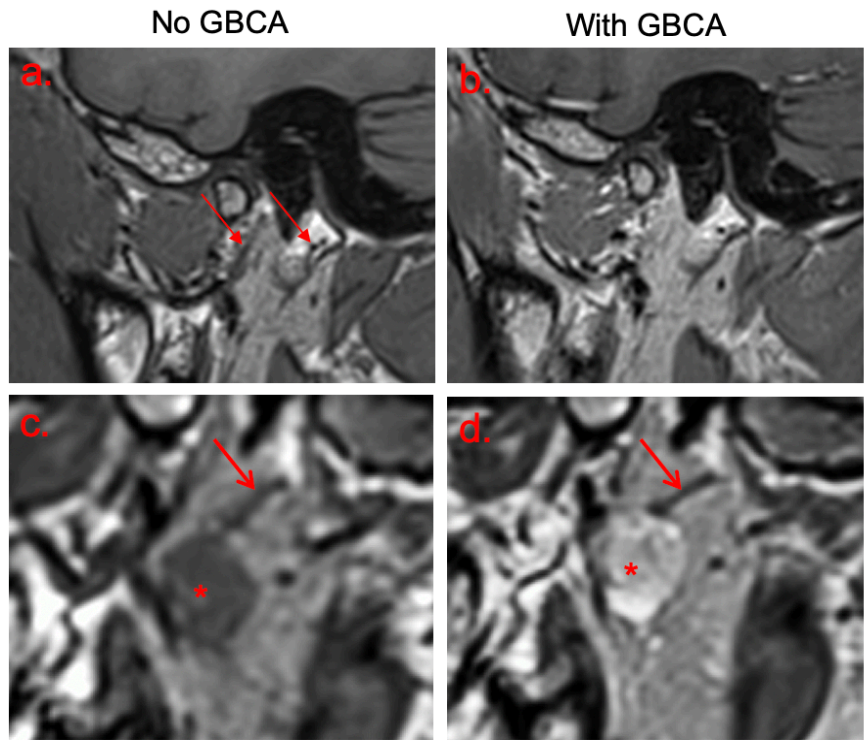

Figure 6 – Method 1 (T1w VFA-TSE) images before and after GBCA. The contrast between the parotid gland and nerve slightly increases on the post-GBCA images (a,b). In addition, some small vessels (closed arrowhead) become less conspicuous on the post-GBCA images, reducing the potential for misidentification. In patients with tumor, the tumor (*) and nerve (open arrowhead) have similar signal intensity on the images without GBCA (c); the tumor significantly increases in intensity on the post-contrast images while the nerve remains low signal (d), meaning that CNR increases. This may be of particular importance where the nerve lies in close proximity to the tumor.